\documentclass[11pt,nofootinbib,showpacs]{revtex4}
\usepackage{amsfonts,amssymb,amsmath,graphicx}
\def\dac{\displaystyle\frac}
\def\[{\left[}
\def\]{\right]}
\def\({\left(}
\def\){\right)}

\begin{document}
\baselineskip7mm

\title{Cosmological dynamics of gravitating hadron matter}

\author{Fabrizio Canfora}
\affiliation{Centro de Estudios Cientificos (CECS), Casilla 1469 Valdivia, Chile}
\affiliation{Universidad Andres Bello, Av. Republica 440, Santiago, Chile}
\author{Alex Giacomini}
\affiliation{Instituto de Ciencias F\'isicas y Matem\'aticas, Universidad Austral de Chile, Valdivia, Chile}
\author{Sergey A. Pavluchenko}
\affiliation{Instituto de Ciencias F\'isicas y Matem\'aticas, Universidad Austral de Chile, Valdivia, Chile}

\begin{abstract}
Anisotropic cosmologies are studied in  the case where the matter source is given by the Skyrme model which is an effective description of low 
energy QCD. The dynamical evolution of the Kantowski-Sachs and Bianchi-I universes are analyzed in depth. In both situations in order for solutions to
exist and at the same time to avoid finite time future singularities, bounds on the value of the cosmological constant and on the values of the Skyrme 
couplings 
must be set. The upper bound on the cosmological constant, which depends also on the initial conditions is closely related to the fact that the baryons
appear below 1 GeV. The lower bound on the cosmological constant and the bounds on the Skyrme couplings are due to the peculiar 
combination of nonlinear terms in the Skyrme model. It is worth to point out  that bounds on the Skyrme  couplings occur in similar fashion both for the 
Kantowski-Sachs and for the Bianchi-I models which are topologically completely different. Our results suggest that this behavior is intrinsic to the coupling of 
the Skyrme field to gravity rather than on a specific cosmological model.
\end{abstract}

\pacs{04.40.-b, 11.10.Lm, 12.39.Dc, 98.80.Cq}
\maketitle

\section{Introduction}

The cosmological constant $\Lambda$ was introduced by Einstein in 1917 to obtain a static universe, but after the discovery of the expansion of the Universe
in 1929 by Hubble, it was dismissed as his ``greatest
blunder''. However the idea of $\Lambda$ had a great comeback
due to recent observational data from supernovae~\cite{perl, perl2, riess},
further supported by Cosmic Microwave Background radiation data~\cite{benoit, benoit3, benoit2, benoit4}, which suggest that the Universe is undergoing
accelerated expansion. Moreover observational data nowadays~\cite{PAR} suggest
that about 70\% of the energy density of the Universe is due to ``dark energy'' and
the cosmological constant $\Lambda$ is the simplest model of dark energy and
it is interpreted as vacuum energy. Other more generic models
consider also a dark energy which is varying in time (for a review see e.g. \cite{quint1, quint2, quint3}). This is usually modeled by scalar fields with some suitable potential, which give rise to a non-linear
theory. More recently also there has been proposed a link between dark energy and the recently discovered Higgs particle \cite{krauss-dent}.

Despite the great success in explaining many features of observational
cosmology, interpretation of the cosmological constant as vacuum energy arising from the Quantum Field Theory (QFT
henceforth) gives rise to a huge problem as its estimated value is $10^{120}$
times larger than what is observed. This remarkable discrepancy,
often called ``the cosmological constant problem''~\cite{L}, is one of the most important
open problems in physics. It is commonly believed that only some
hints from the quantum version of General Relativity can provide us with bounds on $\Lambda$ explaining why it is
non-negative and 120 orders of magnitude smaller than the naive QFT
computation. The main drawback of this point of view is
that the final theory of quantum gravity is not available yet
(although there are strong candidates like string theory~\cite{st_review} and loop quantum gravity~\cite{lqg_review}).
It is however also possible to obtain a bound on the value of the cosmological constant  without invoking any property of
the \textit{still-to-be-discovered} quantum version of General Relativity (see e.g.~\cite{jmartin}). Indeed  it is well known that the Universe from the moment of the Big Bang has 
undergone a series of phase transitions. The last one was the transition from a quark-gluon plasma to a confined phase where matter is colorless. This phase 
transition occurred at an energy scale of approximately 1 GeV. This means that the value of the cosmological constant can not exceed the total energy 
density at the moment of the last phase transition which is of 72 lower than the value obtained from QFT arguments. It worth to point out that this quite 
simple argument to put an upper bound on the cosmological constant has, to the best knowledge of the authors never been stated in literature. Moreover below 
the scale of 1 GeV both General Relativity and the Skyrme model (which is the low energy limit of Quantum Chromo-Dynamics (QCD)\cite{add1, add2}) are 
believed to be perfectly valid. 
Due to these facts it is of great interest to study the properties of cosmological models where the matter field is given by the low energy limit of 
QCD\cite{add1, add2} which is phenomenologically very successfully described by  the Skyrme theory \cite{skyrme, skyrme2, skyrme3} (for a
detailed review see \cite{manton}). The Skyrme
model was historically a generalization of the non-linear Sigma model in order to allow the existence of static
soliton solutions. Remarkably the Skyrme soliton solutions can have excitations with fermionic degrees of freedom also if the basic involved field is bosonic.
This is due to the fact that these soliton solutions can possess a non-trivial winding number. The Skyrme model therefore represents a unified description of
pions and baryons, where the baryon number is given by the winding of the solution. This means that this model is an effective low energy model of QCD making
it one of the most important non-linear theories in nuclear and high energy physics.

The Skyrme theory is characterized by two
coupling constants which, following the notation of~\cite{canfora2}, will be
denoted by $K$ and $\lambda$. Obviously, the values of $K$ and $\lambda$ are
known from experiments; however, for the sake of the present analysis we will
consider both coupling constants as parameters of the theory. Since any
interesting cosmological metric is time-dependent, it is necessary to use a
generalized version of the typical hedgehog {\it ansatz} for the Skyrme field
(introduced in \cite{canfora2, canfora}) which is, at
the same time, compatible with the symmetries of the cosmological backgrounds
of interest and capable to keep track of the non-linear interactions of the
Skyrme theory.

In order to be as general as possible we couple General
Relativity with the Skyrme theory considering two different space-time geometries: one of them is a generalization of the well known flat
Friedmann-Robertson-Walker universe, which is known as Bianchi-I (in order to allow in principle
anisotropies), while the second is Kantowski-Sachs universe with a non-trivial space topology. The energy-momentum
tensor of the Skyrme field suitable for these models can be found
in \cite{canfora2}.

The Kantowski-Sachs metric  has a spatial topology $R\times S_2$. This means that it is natural to use a hedgehog {\it ansatz} for the Skyrme field. Indeed this is
the most studied {\it ansatz} for Skyrmions. Remarkably the equations of motion put a constraint on the existence of solutions in terms of the value of the
cosmological constant and of the Skyrme coupling. Even if  equations of motion can not be integrated an exhaustive numerical analysis shows clearly the
qualitative behavior of the solutions. In order for solutions to exist not only must the cosmological constant be below a certain value that depends on 
the initial conditions but also the couplings of the Skyrme theory must be bounded.  Even in the range of the cosmological and coupling constants where 
solutions exist many solutions present finite time future
singularities. In order to avoid these singularities there arise more constraints on the permitted values of the cosmological constant which can in this case
must be larger than zero. Moreover the existence  of solutions without finite time future singularities also constrain the value of the Skyrme couplings.

The Bianchi-I space-time has topology $R\times R \times R$ and so the hedgehog {\it ansatz} can not be directly used. There exits however a generalization of
this {\it ansatz} as done in \cite{canfora2}. Remarkably also in Bianchi-I, even if with some different details, in order for solutions to exist and 
at the same time to avoid
finite time future singularities, one needs to put constraints on the cosmological constant which must be in this case non-negative and smaller than a 
critical value. Also in this case constraints on the Skyrme couplings arise.

The only assumption that we will use in the present analysis is to require the
existence of solutions which are free of finite-time future singularities (FTFS henceforth). Skyrme-induced FTFS can occur only around 1 GeV
energy scale when the corresponding gravitational effects are strongest. Since the Universe exists in its current form (at a much lower energy scale) we
argue that Skyrme-induced FTFS never occured\footnote{This assumption is very natural itself. On the other end, it can also
be derived with an anthropic argument~\cite{anth}.}.

The structure of the paper is the following: in the second section the Skyrme action is introduced, while in the third section the hedgehog {\it ansatz} is described.
Then the cosmological dynamics of both the Kantowski-Sachs and Bianchi-I universes are analyzed. Finally we draw the conclusions.

\section{The Skyrme action}
The Skyrme model is one of the most famous examples of non-linear field theories as it provides a unified description of
pions and nucleons. This is due to the fact that even if the field is bosonic there exist soliton solutions with non-trivial winding number which possess
fermionic excitations. The winding number is indeed interpreted as baryonic number (see e.g. for a review~\cite{manton}). The Skyrme action can be constructed in the following way: Let be $U$ a
$SU(2)$ valued scalar field. We can the define the quantities:
\begin{equation*}
R_{\mu}^i t_i \equiv R_{\mu}=U^{-1}\nabla_{\mu}U,
\end{equation*}
\begin{equation*}
F_{\mu \nu}=[R_{\mu},R_{\nu}].
\end{equation*}
Here the Latin indices correspond to the group indices and the generators $t_i$ of $SU(2)$ are related to the Pauli matrices by $t_i =-i \sigma_i$.
The Skyrme action is then defined as
\begin{equation*}
S_{Skyrme}=\frac{K}{2}\int d^4 x\sqrt{-g}\mathrm{Tr}\left( \frac{1}{2}R_{\mu}R^{\mu}+ \frac{\lambda}{16}F_{\mu \nu}F^{\mu \nu} \right).
\end{equation*}
The case when $\lambda =0$ is called non linear Sigma Model and the term which multiplies $\lambda$ is called the Skyrme term. Historically the Skyrme term was introduced due to the fact that the non linear Sigma Model does not admit static soliton solutions. The total action for a self gravitating Skyrme field reads
\begin{equation*}
S=\frac{1}{16\pi G}\int d^4x \sqrt{-g}(R-2\Lambda)+S_{Skyrme}.
\end{equation*}
\section{The hedgehog ansatz}
The $SU(2)$ valued scalar field can be parametrized in a standard way
\begin{equation*}
U=\mathbf{I} Y^0 + Y^it_i \; \; \; ; \; \; \; U^{-1}= \mathbf{I} Y^0 - Y^it_i, \label{Ufield}
\end{equation*}
with $Y^0=Y^0(x^{\mu})$ and $Y^i=Y^i(x^{\mu})$ must satisfy $(Y^0)^2+Y_iY^i=1$.\label{Uparametrization}
The most famous and most studied {\it ansatz} for searching solutions to the (non-self gravitating) Skyrme theory is the so called ``hedgehog'' which is obtained by choosing
\begin{equation}
 Y^0 =\cos (\alpha ) \; \; \; ; \; \; \;  Y^i= n^i\sin (\alpha ), \label{hedgehog}
\end{equation}
where $\alpha$ is a radial profile function and $n^i$ is a normal radial vector
\begin{equation}
n^1=\sin \theta \cos \phi \; \; \; ; \; \; \; n^2 = \sin \theta \sin \phi \; \; \; ; \; \; \; n^3=\cos \theta. \label{normalvector}
\end{equation}
It is worth to stress that the hedgehog {\it ansatz} mixes inner degrees with space degrees of freedom as the ``normal vector'' $n^i$ lives in the inner space but is parametrized by space indices. In other words at every space point the vector $n^i$ point radially in the inner space.
This {\it ansatz} which has been studied exhaustively in flat space-time can also be used in curved space-times with spherical symmetry \cite{canfora2}
\begin{equation}
ds^2= g_{AB}dy^A dy^B+ r(y)^2\gamma_{ab}dz^adz^b, \label{sphericalmetric}
\end{equation}
where $g_{AB}$ and $y^A$ are metric and coordinates of a two dimensional Lorentzian manifold, $r(y)$ is a warp function and $\gamma_{ab}$ and $z^a$ are the metric and coordinates on the two sphere $\gamma_{ab}dz^adz^b= d\theta^2+\sin ^2\theta d\phi$. In this case the hedgehog {\it ansatz} remains the same and in general the profile function $\alpha= \alpha (y)$ is a function of both coordinates of the two dimensional Lorentzian manifold.
The stress tensor of the Skyrme field for (\ref{hedgehog})--(\ref{sphericalmetric}) then be written as
\begin{equation}
\begin{array}{l}
T_{\mu \nu}dx^{\mu}dx^{\nu}=K \left[ \(1+2\lambda r^{-2}\sin ^2 \alpha\)\(\partial_A \alpha \partial_B \alpha
-\frac{1}{2}g_{AB} \partial_C \alpha \partial^C \alpha\) - g_{AB} r^{-2}\sin ^2\alpha \times \right. \\ \\ \left. \times \(1+\frac{\lambda}{2}r^{-2}\sin ^2\alpha\) \right]dy^Ady^B
-\frac{1}{2}K\(\partial_C \alpha \partial^C \alpha-\lambda r^{-4}\sin ^4 \alpha\)r^2\gamma_{ab}dz^adz^b.
\end{array}\label{hedgehogtmunu}
\end{equation}

\section{Kantowski-Sachs space}
In the context of cosmology, in order to use the hedgehog {\it ansatz} it is necessary to have a two dimensional sphere as a sub-manifold. Therefore the natural
choice is the Kantowski-Sachs metric.

The Kantoski-Sachs space-time describes a homogeneous but not isotropic universe and has a topology  $R\times S_2$. This space-time is  very remarkable
because it describes the only homogeneous cosmology which is not included in the Bianchi classification. This is due to the fact that it does not have a
three dimensional isometry subgroup which is simply transitive. The metric can be written as

\begin{equation*}
\begin{array}{l}
ds^2 = -dt^2 + a(t)^2 dr^2 + b(t)^2 (d\theta^2 + \sin^2\theta d\phi^2)
\end{array}\label{KS_metrics}
\end{equation*}
where $a(t)$ and $b(t)$ are two functions which are determined by the Einstein field equations.

\subsection{Equations of motion}
The equation of motion for the Skyrme field reads \cite{canfora2}
\begin{equation}
\begin{array}{l}
\( \ddot\alpha + H_a\dot \alpha\) \(1+\dac{2\lambda}{b^2} \sin^2\alpha \) + \dac{2\dot b\dot\alpha}{b} + \dac{\sin(2\alpha)}{b^2}
\[1+\lambda\( \dot\alpha^2 + \dac{\sin^2\alpha}{b^2}\)\] = 0,
\end{array}\label{skyrme}
\end{equation}

\noindent where\footnote{Since $a(t)$ corresponds to $R^1$ part of the metric, there is no spatial curvature associated
with it and so, similarly to the spatially flat Friedmann cosmology, corresponding scale factor holds no physical meaning and the equations
should be rewritten in terms of Hubble parameter $H_a\equiv \dot a/a$.} $H_a\equiv \dot a/a$,
$a\equiv a(t)$, $b\equiv b(t)$, $\alpha \equiv \alpha(t)$ and dot above represents the derivative with respect to time $t$.

For the Kantowski-Sachs metric it is straightforward to see that the stress-energy tensor (\ref{hedgehogtmunu}) reads:

\begin{equation}
\begin{array}{l}
T_{00} = K\[ \dac{\dot\alpha^2}{2} \(1+\dac{2\lambda}{b^2} \sin^2\alpha \) + \dac{\sin^2 \alpha}{b^2} \( 1+ \dac{\lambda\sin^2 \alpha}{2b^2}\)\], \\ \\
T_{11} = K a^2 \[ \dac{\dot\alpha^2}{2} \(1+\dac{2\lambda}{b^2} \sin^2\alpha \) - \dac{\sin^2 \alpha}{b^2} \( 1+ \dac{\lambda\sin^2 \alpha}{2b^2}\)\], \\ \\
T_{22} = \dac{Kb^2}{2} \( \dot\alpha^2 + \dac{\lambda\sin^4 \alpha}{b^4}\), \\ \\
T_{33} = \sin^2 \Theta T_{22};
\end{array}\label{set}
\end{equation}

\noindent while the components of the Einstein tensor are:

\begin{equation}
\begin{array}{l}
G_{00} =  \dac{\dot b^2 + 1}{b^2} + 2H_a \dac{\dot b}{b}, \\ \\
G_{11} = -a^2\( \dac{\dot b^2 + 1}{b^2} + 2\dac{\ddot b}{b} \), \\ \\
G_{22} = -\dac{b}{a} \( \ddot a b + \dot a \dot b + a \ddot b \), \\ \\
G_{33} = \sin^2 \Theta G_{22},
\end{array}\label{et}
\end{equation}

\noindent which give us three dynamical equations (Skyrme as well as 11 and 22 components of Einstein equations $T_{ij} = G_{ij}$ where we
put $8\pi G/3 \equiv 1$) and constraint equation (00 component of Einstein equations); for obvious reasons we do not consider 33 component
of Einstein equations.

\subsection{Skyrme case}

In this section we study behavior of the model with just Skyrme term as a source, i.e. described by (\ref{skyrme})--(\ref{et}). We study the
system numerically: we choose initial values for $\alpha$, $b$, $\dot b$, and $H$ and calculate initial value for $\dot \alpha$ from the
constraint ($T_{00}=G_{00}$) equation. Then we integrate dynamical equations (11 and 22 components of Einstein equations as well as Skyrme
field equation) in both directions in time to see the whole time evolution which corresponds to particular set of initial conditions.

\begin{figure}
\includegraphics[width=1.0\textwidth, angle=0, bb=0 0 648 216]{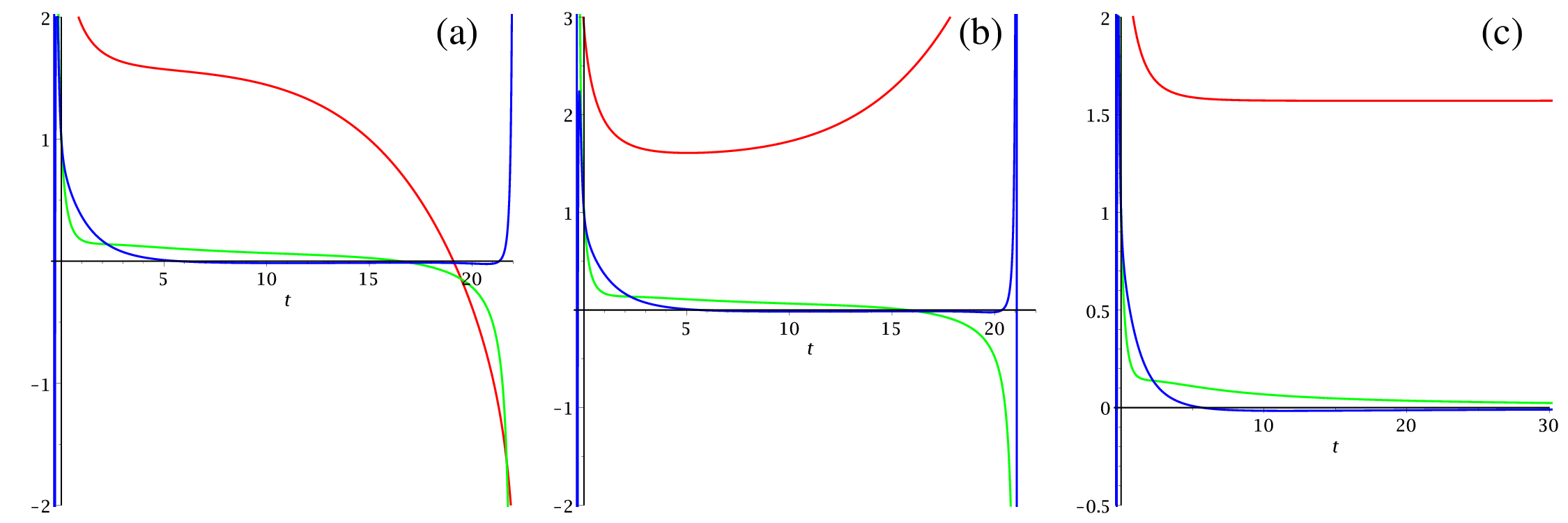}
\caption{Typical behavior of the system with only Skyrme term as a source. Red line corresponds to $\alpha$, green -- to $\dot b/b$, and
blue -- to $H$. Panels (a) and (b) reflect typical singular at late time behavior while (c) corresponds to unstable regime with
$\alpha\to \pi/2 + \pi n, n\in \mathbb{Z}$ (see text for details).}\label{fig1}
\end{figure}

The past behavior is singular for all initial conditions, while future is a bit different. Most of the initial conditions lead to
finite-time future singularity with $\alpha\to\pm\infty$, as presented in Fig. \ref{fig1}(a) ($\alpha\to -\infty$) and  Fig. \ref{fig1}(b)
($\alpha\to +\infty$). Areas with different sign of $\alpha$ at future singularity are separated by unstable trajectories presented in
Fig. \ref{fig1}(c) where \mbox{$\alpha\to \pi/2 + \pi n, n\in \mathbb{Z}$}, so that the final values for $\alpha$ for two consecutive separation lines 
\linebreak 
differ in $\pi$.

\subsection{$\Lambda$-term case}

\begin{figure}
\includegraphics[width=0.67\textwidth, angle=0]{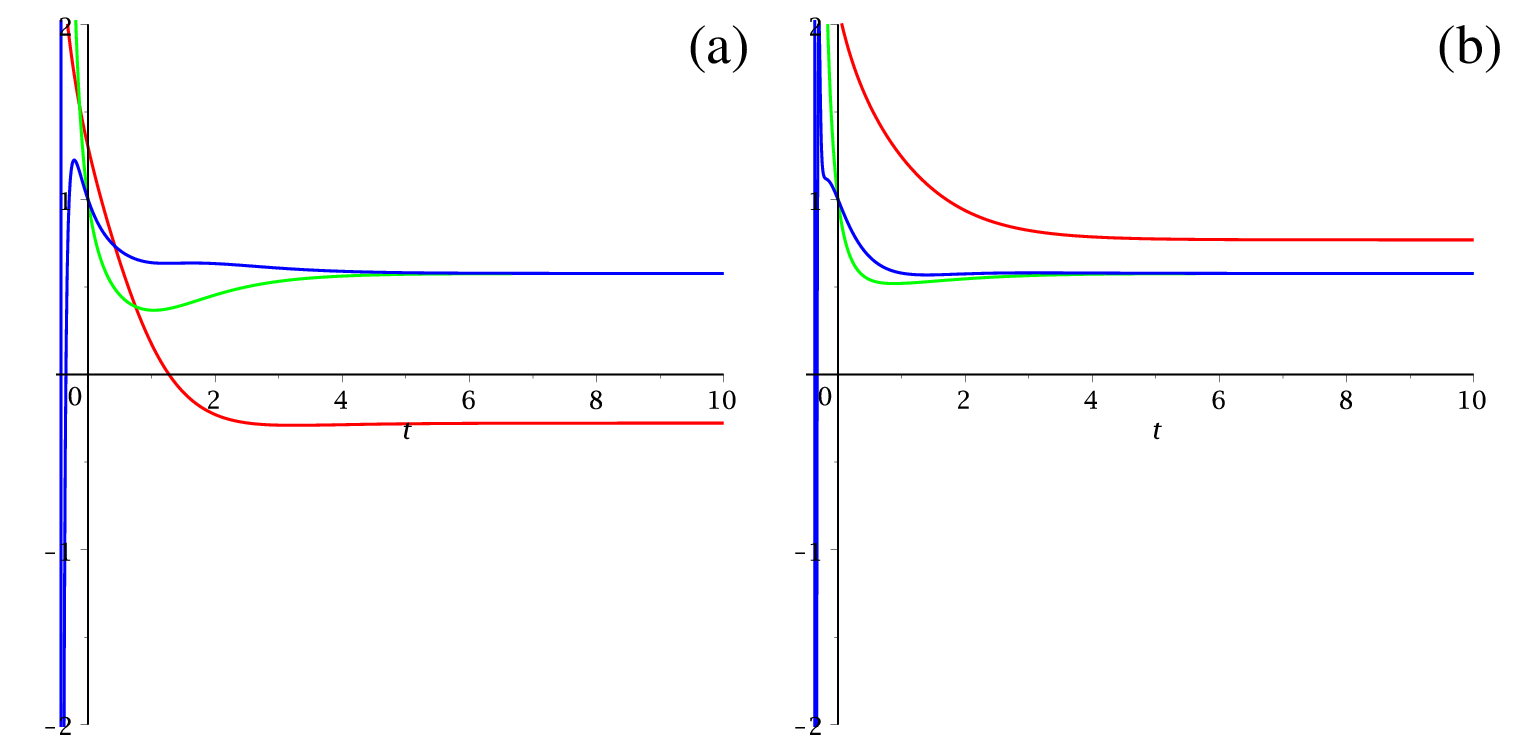}
\caption{Typical behavior of the system with both Skyrme and $\Lambda$-term as a source. Notations are the same as in Fig. \ref{fig1}.
Panels (a) and (b) have slightly different initial conditions and demonstrate that final value for $\alpha$ is no longer quantized compared
to Skyrme case (see text for details).}\label{fig2}
\end{figure}

In this section we added $\Lambda$-term as an additional source to the system under consideration; the equations (\ref{skyrme})--(\ref{et})
are modified in a similar way as they modified in standard cosmology when one adds $\Lambda$-term. The behavior changes drastically in presence
of nonzero positive $\Lambda$-term and typical is shown in Fig. \ref{fig2}. The main difference is that we no longer have finite-time future
singularity 
regardless of the initial conditions. The other difference is that the value for $\alpha$ at infinity is no longer quantized compared to the
``pure'' Skyrme case. The explanation for this we see as follows: with the Skyrme being ordinary matter, its energy density, no matter now high it was,
is eventually decreasing with the expansion whilst $\Lambda$-term energy-density remains constant. At some point the Skyrme contribution
becomes subdominant and the expansion dominated by Skyrme changes to de Sitter phase. Soon after the value for the Skyrme field $\alpha$
``freezes'' near its current value which could be any -- we demonstrated it in Fig. \ref{fig2}.

\begin{figure}
\includegraphics[width=1.0\textwidth, angle=0, bb=0 0 396 140]{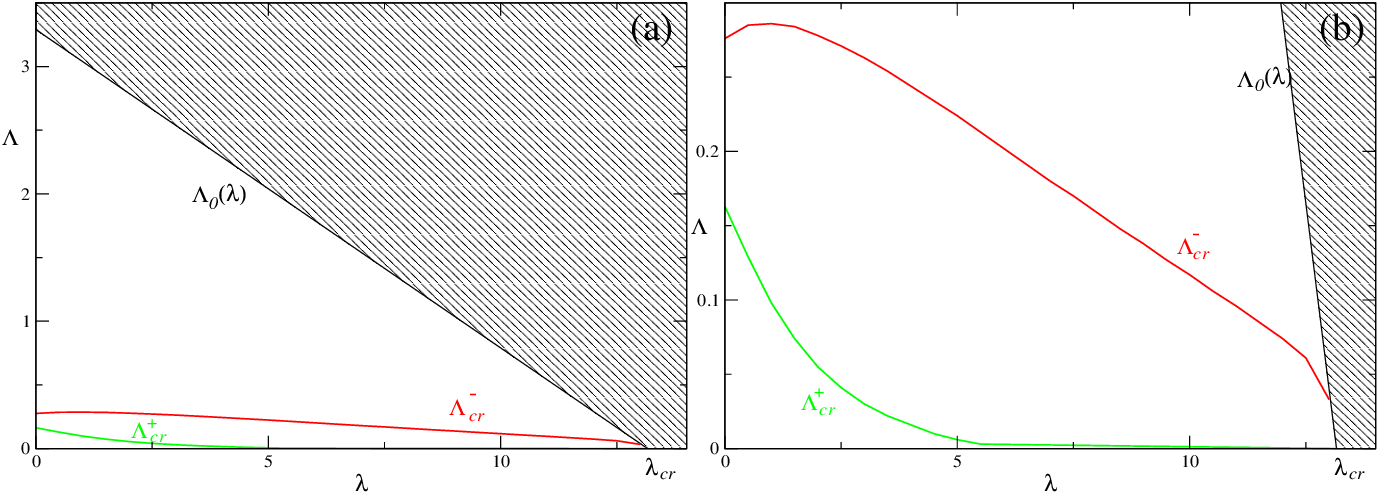}
\caption{The areas in the ($\lambda$, $\Lambda$) space which lead to isotropisation ($\Lambda > \Lambda_{cr}^\pm$) or to finite-time
future singularity ($\Lambda < \Lambda_{cr}^\pm$); supscript ``$\pm$'' corresponds to positive or negative branch of $\dot\alpha$ as solved from (\ref{00});
shaded
region depicts ``forbidden'' region of ($\lambda$, $\Lambda$) combinations
(see text for details). Panel (a) presents ``large-scale'' picture with the entire allowed region in the first quadrant depicted while (b) panel
focuses on the smaller region with detailed $\Lambda_{cr}^\pm$ curves.
Numerical values for $\Lambda$ and $\lambda$ are subject to the initial conditions.
}\label{fig3}
\end{figure}

There are two cases when in presence of nonzero $\Lambda$-term we still have finite-time future singularity. First of them is when $\Lambda$-term
is negative -- in that case we always have finite-time future singularity. The second case is when $\Lambda$-term is positive, but its value is
low enough for future singularity to occurs before $\Lambda$-term
becomes dominant; this case is represented in Fig. \ref{fig3}. There we presented $\Lambda_{cr}(\lambda)$ curves for for two branches for initial
value of $\dot\alpha$. Indeed, 00-component of Einstein equation

\begin{equation}
\begin{array}{l}
K\[ \dac{\dot\alpha^2}{2} \(1+\dac{2\lambda}{b^2} \sin^2\alpha \) + \dac{\sin^2 \alpha}{b^2} \( 1+ \dac{\lambda\sin^2 \alpha}{b^2}\)\] =
\dac{\dot b^2 + 1}{b^2} + 2H_a \dac{\dot b}{b} - \Lambda,
\end{array}\label{00}
\end{equation}

\noindent being solved with respect to $\dot\alpha^2$, leads to two branches depending on the sign (``+'' or ``--'') for $\dot\alpha$.
So in Fig. \ref{fig3} for these two branches we plot the value for $\Lambda_{cr}^\pm$ so that if for $\Lambda > \Lambda_{cr}^\pm$ we always have
isotropisation and for $\Lambda < \Lambda_{cr}^\pm$ we always have finite-time future singularity for a given set of initial conditions
($\alpha_0$, $b_0$, $\dot b_0$, and $H_0$) and given value for $K$. Despite the fact that actual values
for $\Lambda_{cr}(\lambda)$ depend on initial conditions and $K$, general behavior remains the same with one exemption.

This case is presented in Fig. \ref{fig3} as a shaded region. And this region corresponds to
``forbidden'' region of
$(\Lambda,\,\lambda)$ values. Indeed, if one solves (\ref{00}) with $\dot\alpha=0$ with respect to, say, $\Lambda$, one would get

\begin{equation}
\begin{array}{l}
\Lambda_0 \equiv \Lambda = A - B\lambda,~\mbox{where}~A=\dac{4\dot b b^3 H + 2b^2 \dot b^2 + 2b^2 - 2Kb^2\sin^2\alpha}{2b^4},
~B=\dac{K\sin^4\alpha}{2b^4}.
\end{array}\label{lambda0}
\end{equation}

\noindent One can see that $B \geqslant 0$ always while $A$ potentially could have both signs; in Fig. \ref{fig3} initial conditions correspond to $A > 0$.
One can also easily check that for $\Lambda > \Lambda_0$ the value for $\dot\alpha_0^2$ from (\ref{00}) becomes negative which means that one cannot choose
$\Lambda > \Lambda_0$. In conclusion: for any $\lambda < \lambda_{cr}$ there exists $\Lambda_0$ defined from (\ref{lambda0}) so that $\Lambda > \Lambda_0$ are
forbidden from energy budget. And $\lambda_{cr}$ is defined from (\ref{lambda0}) where the line crosses $\Lambda=0$ -- it would be $\lambda_{cr} = B/A$
in terms of (\ref{lambda0}). So for $\lambda > \lambda_{cr}$ we must have $\Lambda < 0$ in order for any solution to exist; needless to say, in that case
we always have finite-time future singularity.

One can also see that $A < \infty$ -- indeed, in order for $A$ to diverge one needs either diverge one of the terms in numerator (so to diverge $H$, $b$, or $\dot b$
which means divergence of $G_{ij}$ and so physical singularity) or put denominator to zero -- so that to put $b=0$ which also means singularity. So that
in all regular points $A$ is finite and so limit on $\Lambda$ is also finte.

We mentioned that $A$ could be negative -- in that case the interception of the $\Lambda_0(\lambda)$ with $\lambda = 0$ would occur at negative values
of $\Lambda$ making {\it all} possible values for both $\Lambda$ and $\lambda$ lead to finite-time future singularity. That happens if $K > K_{cr}$ which
is defined from $A=0$ from~(\ref{lambda0}):

\begin{equation}
\begin{array}{l}
K_{cr} = \dac{1 + \dot b^2 + 2bH \dot b}{\sin^2\alpha}.
\end{array}\label{kcr}
\end{equation}

\subsection{Discussion}

One can see several differences between the regimes with no finite-time future singularity with and without $\Lambda$-term. First is the
abundance of the regime -- in pure Skyrme case this regime is unstable and has zeroth measure on the initial conditions space whilst in
$\Lambda$-case it has nonzero measure.

The second difference is the ``final'' value for $\alpha$ -- in pure Skyrme case it is quantized \mbox{$\alpha_\infty = \pi/2 + \pi n, n \in \mathbb{Z}$}
whilst
in the $\Lambda$-case it can take any value. This difference is due to the fact that in pure Skyrme term the dynamics is driven by the Skyrme
``till the end'', while in the $\Lambda$-case the cosmological term eventually (if $\Lambda > \Lambda_0^\pm$) become dominant resulting in
changing the dynamics to de Sitter stage. When it happenes, the value for the Skyrme ``freezes'' and it can happen at any value.

And the last, but not least difference we would like to point an attention to, is the dinamical behavior of the metric functions. One can see
that in the Skyrme case $H$ quickly reaches near-zero value and remains there, whilst $\dot b/b$ (it has the same dimensionality as $H$) remains
positive and slowly decreasing. On the other hand, in the $\Lambda$-case we have $\dot b/b = H$ at late stages, which is typical for
$\Lambda$-dominated cosmologies. So very carefully we can say that in the $\Lambda$-case the Universe isotropize in the de Sitter sense, i.e.
in the de Sitter sense the Hubble parameter plays a role of exponent for exponential de Sitter expansion so when these exponents coinside for
anisotropic model, they call it isotropic. But formally one cannot ``isotropize''  this model -- topologically Kantowski-Sacks represented by
$R^1 \times S^2$ which is anisotropic by its topology and under no dynamics it cannot turn into, say,  $R^3$ which could be formally
isotropize. It will not hurt adding that, as it was already mentioned, the $R^1$ part has no spatial curvature and so the associated scale factor
has no physical meaning, and, keeping in mind that formally isotropisation is the situation when all relevant scale factors are equal, this
way of isotropisation determination cannot be applied to the model under consideration.

We discovered that there exist critical values for all three parameters of the system -- $\Lambda$, $\lambda$ and $K$. First of all, there exists
line $\Lambda_0(\lambda)$, defined by (\ref{lambda0}), which separates allowed ($\Lambda < \Lambda_0$) and forbidden ($\Lambda > \Lambda_0$)
regions on the $(\Lambda,\,\lambda)$ space. If $A > 0$ which happening if $K < K_{cr}$ defined from (\ref{kcr}), then inside the allowed region in the
first quadrant there exist $\Lambda_{cr}^\pm$ lines (see Fig.\ref{fig3}) which separates regions which leads to isotropisation ($\Lambda > \Lambda_{cr}$)
from regions with finite-time future singularity ($\Lambda < \Lambda_{cr}$). Allowed region in first quadrant is bounded by $\lambda_{cr} = B/A$ in terms
of (\ref{lambda0}); so that for $\lambda > \lambda_{cr}$ we always need $\Lambda < 0$ in order to have any solution, and these solution always have
finite-time future singularity. The same -- necessity of $\Lambda < 0$ and finite-time future singularity -- is true if we have $A < 0$ and so $K > K_{cr}$.

So to summarize the predicament for having isotropisarion in this model (given that otherwise we have finite-time future singularity, isotropisation
seems much more physical if we want to consider the model): we need $\Lambda_{cr}^\pm < \Lambda < \Lambda_0$,
$\lambda < \lambda_{cr}$ and $K < K_{cr}$. So that in this particular model just assumption of the physical behavior allows us to put constraints on the
coupling constants and $\Lambda$. Let us note that if we ``turn off'' Skyrme term (by putting $\lambda=0$) with nonlinear Sigma Model still intact ($K\ne 0$)
the constrain on $\lambda$ is obviously lifted but constraint on $K$ remains; in that case the dependence of $\Lambda$ on $\lambda$ in Fig. \ref{fig3} is
replaced with dependence on $K$ like in Bianchi-I case (see Fig. \ref{fig5}). From (\ref{lambda0}) one can see that $\Lambda_0$ depends on both $\lambda$
and $K$ linearly and Fig \ref{fig3} is $K={\rm const}$ slice while Fig. \ref{fig5} could be considered as a reference to $\lambda={\rm const}$ slice.

We summarized all possible regimes in Table \ref{table_ks}.

\begin{table}
\caption{Summary of the regimes in Kantowski-Sachs Skyrme cosmology}
\centering
\label{table_ks}
\begin{tabular}{|c|l|}

\hline

$\Lambda<0$ & only FTFS \\ \hline
$0\leqslant\Lambda<\Lambda_{cr}^\pm$ & FTFS as well as unstable regime with $\alpha\to\pi/2 + \pi n, n\in\mathbb{Z}$ \\ \hline
$\Lambda_{cr}^\pm < \Lambda < \Lambda_0$ & isotropisation \\ \hline
$\Lambda_0 < \Lambda$ & forbidden from energy budget \\ \hline

\end{tabular}
\end{table}

\section{Bianchi-I space}

We have already mentioned that the Kantowski-Sachs cosmology does not fit in the Bianchi classification. On the other hand one of the most interesting
Bianchi type cosmologies is the Bianchi-I cosmology which has topology $R \times R \times R$ and it is totally non-isotropic as it has a different expansion
factor in every spatial direction. The metric can be written as

\begin{equation*}
\begin{array}{l}
ds^2 = -dt^2 + a(t)^2dx^2 + b(t)^2 dy^2 + c(t)^2 dz^2.
\end{array}\label{BI_metrics}
\end{equation*}

Of course in this case we can not use the hedgehog {\it ansatz} (\ref{hedgehog}) as this space-time has not a 2-sphere as a sub-manifold.
It was shown in \cite{canfora2} that the hedgehog {\it ansatz} can be generalized also to non-spherically symmetric space-time by using as {\it ansatz} again
\begin{equation}
 Y^0 =\cos (\alpha ) \; \; \; ; \; \; \;  Y^i= n^i\sin (\alpha ), \label{generalizednormalvector_pre1}
\end{equation}
but now the vector $n^i$ has a different form namely
\begin{equation}
n^1= \cos \Theta \; \; \; ; \; \; \; n^2 = \sin \Theta  \; \; \; ; \; \; \; n^3=0, \label{generalizednormalvector}
\end{equation}
where $\Theta$ is some scalar function.
In this case the energy momentum tensor reads ((4.15) from~\cite{canfora2})
\begin{equation}
\begin{array}{l}
T_{\mu \nu}= K \left[ \left. \nabla _{\mu} \alpha \nabla _{\nu} \alpha +\sin ^2\alpha \nabla _{\mu} \Theta \nabla _{\nu} \Theta + \lambda \sin^2 \alpha\right. \right. \\ \\
\left. \times \left( (\nabla \Theta)^2\nabla _{\mu} \alpha \nabla _{\nu} \alpha + (\nabla \alpha)^2\nabla _{\mu} \Theta \nabla _{\nu} \Theta\right) \right. \\ \\
\left.-\frac{1}{2}g_{\mu \nu}\left(  (\nabla \alpha)^2 +\sin^2 \alpha (\nabla \Theta)^2+\lambda \sin^2 \alpha(\nabla \alpha)^2(\nabla \Theta)^2\right) \right]
\end{array}\label{gent}
\end{equation}
Now we have to make an {\it ansatz} for the scalar function $\Theta$ which respects the symmetry of the Bianchi-I space-time. A possible choice is
\begin{equation}
\Theta= I_1x +I_2y +I_3z \label{Theta}.
\end{equation}
\subsection{Equations of motion}

The equations of   of motion for the Skyrme field are:

\begin{equation*}
\begin{array}{l}
\ddot \alpha \[ 1 + \lambda\sin^2\alpha \( \dac{I_1^2}{a^2} + \dac{I_2^2}{b^2} + \dac{I_3^2}{c^2}  \) \] +
\dot\alpha^2\lambda\sin\alpha\cos\alpha \( \dac{I_1^2}{a^2} + \dac{I_2^2}{b^2} + \dac{I_3^2}{c^2} \) + \\ \\ +
\dot\alpha \[ H_a + H_b +H_c + \lambda\sin^2\alpha \( \dac{I_1^2}{a^2} \( H_b+H_c-H_a \) + \dac{I_2^2}{b^2} \( H_a+H_c-H_b \) +
\dac{I_3^2}{c^2} \( H_a+H_b-H_c \)    \)   \] + \\ \\ +
\sin\alpha\cos\alpha \( \dac{I_1^2}{a^2} + \dac{I_2^2}{b^2} + \dac{I_3^2}{c^2}  \) = 0,
\end{array}\label{skyrme_BI}
\end{equation*}

\noindent where we once again for simplicity put $a\equiv a(t)$, $b\equiv b(t)$, $c\equiv c(t)$, $\alpha \equiv \alpha(t)$, $H_a = \dot a/a$,
$H_b = \dot b/b$, $H_c = \dot c/c$ and dot represents derivative with respect to time and we used (\ref{Theta}) as an {\it ansatz} for $\Theta$.

Since Bianchi-I metrics is spatially flat, we rewrite the equations in terms of corresponding Hubble parameters (e.g. $H_a\equiv \dot a/a$) where it is
convinient. But our source introduce ``curvature-like'' terms (like $I_1^2/a^2$) so that unlike ``classical'' Bianchi-I cosmological models we cannot
rewrite equations of motion solely in terms of Hubble parameters.

The components of the stress-energy tensor using (\ref{Theta}) in (\ref{gent}) read

\begin{equation}
\begin{array}{l}
T_{00} = \dac{K}{2}\[  \dot\alpha^2 + \sin^2\alpha \( 1+ \lambda\dot\alpha^2\) \( \dac{I_1^2}{a^2} + \dac{I_2^2}{b^2} + \dac{I_3^2}{c^2}  \)  \], \\ \\
T_{11} = a^2 \dac{K}{2}\[  \dot\alpha^2 + \sin^2\alpha \( 1- \lambda\dot\alpha^2\) \( \dac{I_1^2}{a^2} - \dac{I_2^2}{b^2} - \dac{I_3^2}{c^2}  \)  \], \\ \\
T_{22} = b^2 \dac{K}{2}\[  \dot\alpha^2 + \sin^2\alpha \( 1- \lambda\dot\alpha^2\) \( -\dac{I_1^2}{a^2} + \dac{I_2^2}{b^2} - \dac{I_3^2}{c^2}  \)  \], \\ \\
T_{33} = c^2 \dac{K}{2}\[  \dot\alpha^2 + \sin^2\alpha \( 1- \lambda\dot\alpha^2\) \( -\dac{I_1^2}{a^2} - \dac{I_2^2}{b^2} + \dac{I_3^2}{c^2}  \)  \], \\ \\
T_{ij} = K \sin^2\alpha I_i I_j \( 1 - \lambda\dot\alpha^2\).
\end{array}\label{set_bi}
\end{equation}

The components of the Einstein tensor are:

\begin{equation}
\begin{array}{l}
G_{00} =  H_aH_b + H_aH_c + H_bH_c, \\ \\
G_{11} = -a^2\( \dot H_b + \dot H_c + H_b^2 + H_c^2 + H_bH_c \), \\ \\
G_{22} = -b^2\( \dot H_a + \dot H_c + H_a^2 + H_c^2 + H_aH_c \), \\ \\
G_{33} = -c^2\( \dot H_a + \dot H_b + H_a^2 + H_b^2 + H_aH_b \).
\end{array}\label{et_bi}
\end{equation}

One can immediately see from (\ref{set_bi}) and (\ref{et_bi}) that for spatial $i\ne j$ $G_{ij}\equiv 0$ while generally $T_{ij}\ne 0$. So that we have to
require $T_{ij}|_{i\ne j} = 0$ to fulfill Einstein equations. This could be done in three ways: we either fine-tune $\alpha(t)$ to fulfill
$( 1 - \lambda\dot\alpha^2) = 0$, or put two out of three $I_i$ to zero (to fulfill all three $T_{ij} = 0$) or fine-tune $\alpha(t)$ so that
$\sin^2\alpha\equiv 0$. Last way imply $\alpha = \pi n,~n\in\mathbb{Z}$ which in turn produce trivial results -- in that case $T_{\mu\nu}$ nullify
identically. The first way imply $\alpha(t) = \alpha_0 \pm t/\sqrt{\lambda}$, but this {\it ansatz} is inconsistent with system of diagonal equations.
So all that remains is to nullify two out of three $I_i$. 

\subsection{Cosmological dynamics}

\begin{figure}
\includegraphics[width=1.0\textwidth, angle=0, bb=0 0 575 287]{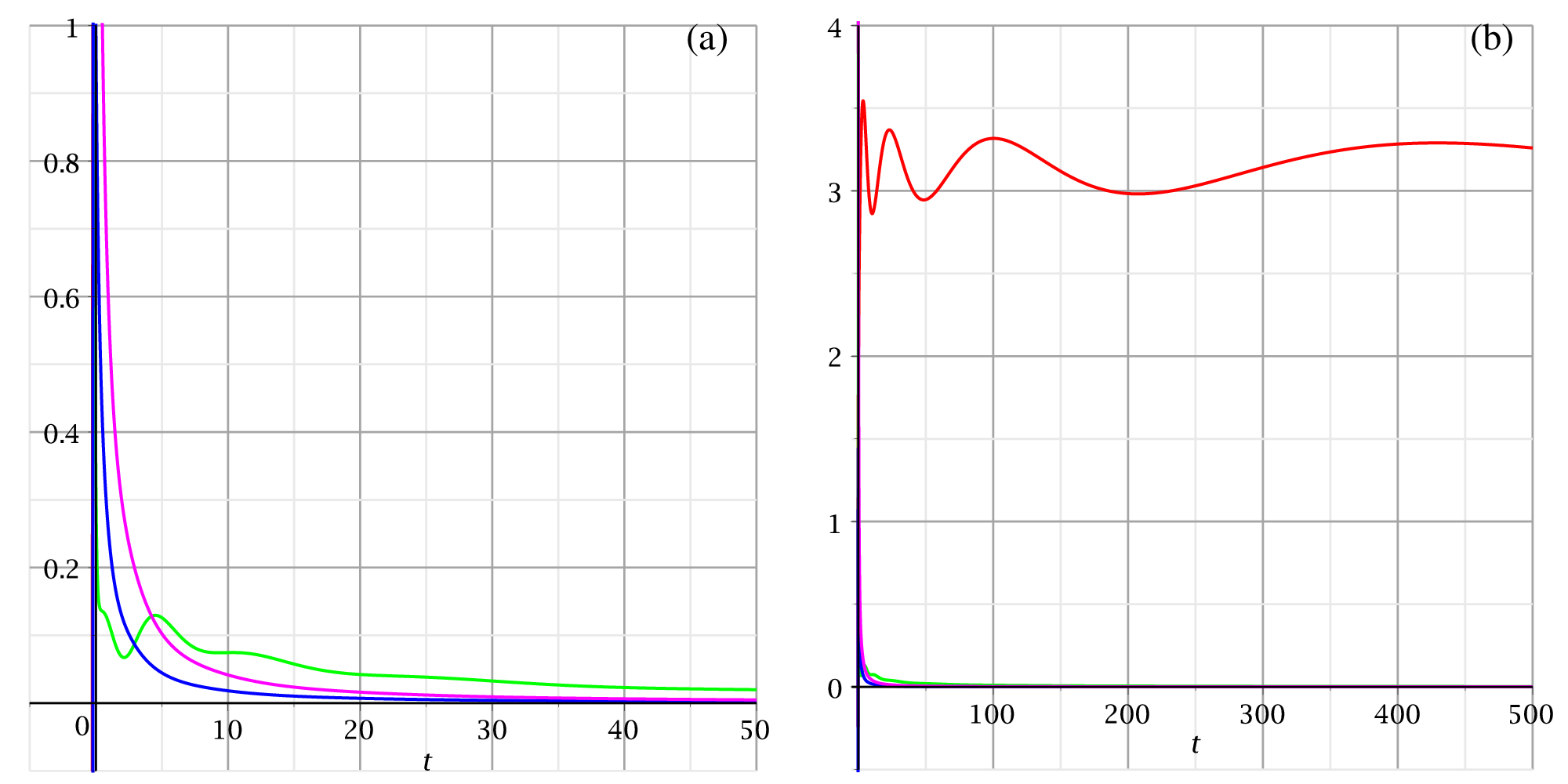}
\caption{The dynamics of the Bianchi-I model with just Skyrme as a source. In (a) panel we presented the behavior of the Hubble parameters, associated with
different scale factors: green corresponds to $H_a$, blue to $H_b$ and magenta to $H_c$. In (b) panel we extended (a) panel so to demonstrate the behavior
of the Skyrme (red). See also text for details.
}\label{fig4}
\end{figure}

The dynamics of the Bianchi-I model is very different from Kantowski-Sachs and resembles that of classical cosmology. In Skyrme case (with $\Lambda = 0$) the
dynamics is presented in Fig.~\ref{fig4}. One can see that generally all three Hubble parameters tend to zero, but one cannot miss that the green curve
(upper one on large $t$) which corresponds to $H_a$ behaves a bit different -- that is due to the fact that in this particular example we chose $I_1\ne 0$
(and this is the
direction $H_a$ is associated with) while $I_2=I_3=0$.
Here one can see that the choice of  which of $I_i$ is nonzero slightly affects the dynamics.
As of the behavior of the Skyrme field, in this particular example it has $\alpha_\infty = \pi$ while generally it is
$\alpha_\infty = \pi n,~n\in\mathbb{Z}$ (including zero).

In case with positive cosmological constant we always have isotropisation -- just as in ``classical cosmology'' case so that the evolution curves resemble
those in Fig. \ref{fig2} with the difference that now we have three Hubble parameters. Finally in case of negative cosmological constant we always have
finite-time future singularity.

\begin{figure}
\centering
\includegraphics[width=0.5\textwidth, angle=0, bb=0 0 655 474]{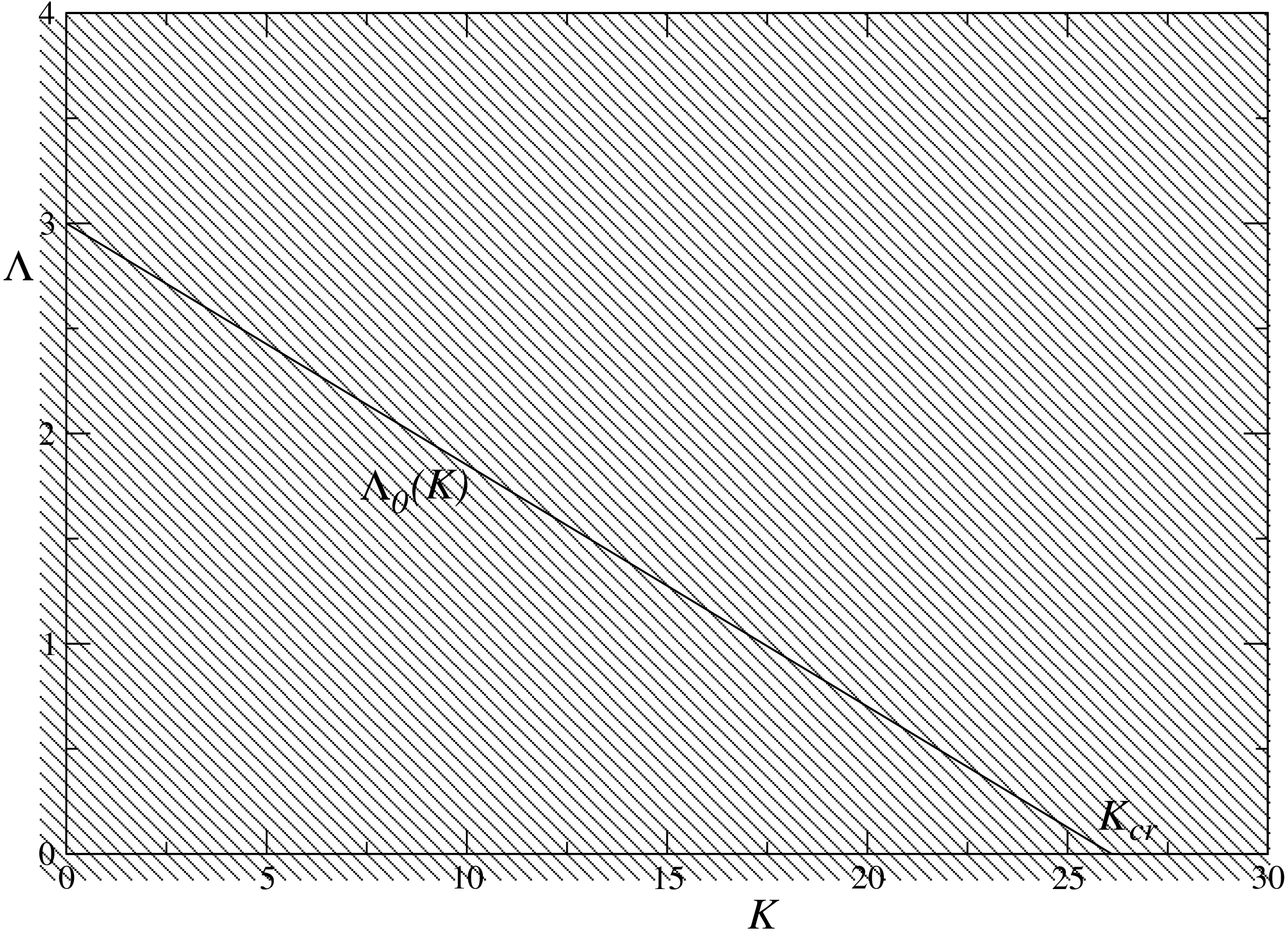}
\caption{The analogue of Fig. \ref{fig3} for Bianchi-I case. We plotted the line $\Lambda_0(K)$ which separates the ``forbidden'' (dashed; above the line)
region from ``allowed'' (see text for details). Numerical values for $\Lambda$ and $K$ are subject to the initial conditions.
}\label{fig5}
\end{figure}

But there is a similarity between the dynamics of the Kantowski-Sachs and Bianchi-I and it lies in the existence of the ``allowed'' and ``forbidden''
regions on the parameters space. Indeed, if we repeat the derivation of these regions from previous section, we get that $\Lambda \leqslant \Lambda_0$:

\begin{equation}
\begin{array}{l}
\Lambda_0 \equiv H_aH_b + H_aH_c + H_bH_c - \dac{K\sin^2\alpha}{2a^2},
\end{array}\label{lambda0_bi}
\end{equation}

\noindent and $\Lambda > \Lambda_0$ leads to $\dot\alpha^2 < 0$ and so is forbidden. Unlike Kantowski-Sachs case, though, in Bianchi-I case $\Lambda_0$
depends on the initial conditions and does not depend on $\lambda$, but the dependence on $K$ remains so the analogue of Fig. \ref{fig3} would be
the Fig. \ref{fig5}.
And similar to the previous section we can see that appropriate choice of $K$
can set $\Lambda_0$ negative: indeed, for $K > K_{cr}$ with

\begin{equation}
\begin{array}{l}
K_{cr} = \dac{2a}{bc\sin^2\alpha} \( a\dot b\dot c + \dot a b\dot c + \dot a\dot b c \),
\end{array}\label{kcr_bi}
\end{equation}

\noindent we have $\Lambda_0 < 0$ and so that to have a solution we require $\Lambda < 0$ (so the case with just Skyrme and $\Lambda = 0$ is prohibited). And this
solution contains finite-time future singularity.

\subsection{Discussion}

In this section we preformed the analysis of the Bianchi-I cosmological model in presence of Skyrme with or without cosmological constant. We find that the
dynamics
of this model is close to ``classical'' Bianchi-I in the sense that the dynamical regimes are the same. But despite that there are differences caused
by the choice of the source -- the Skyrme. These differences lead to possibility to set constraints on both $\Lambda$ and $K$ in order to different
regimes to take place. Additionally Skyrme introduces terms in stress-energy tensor that scale as a curvature so that unlike ``classical'' Bianchi-I
cosmology (with e.g. perfect fluid or scalar fields) the equations cannot be rewritten in terms of Hubble parameters. 

We summarize all possible regimes in Table \ref{table_bi}.

\begin{table}
\caption{Summary of the regimes in Bianchi-I Skyrme cosmology}
\centering
\label{table_bi}
\begin{tabular}{|c|l|}

\hline

$\Lambda<0$ & only FTFS \\ \hline
$\Lambda = 0$ & stable regime depicted in Fig. \ref{fig4} \\ \hline
$0<\Lambda<\Lambda_0$ & isotropisation \\ \hline
$\Lambda_0<\Lambda$ & forbidden from energy budget \\ \hline

\end{tabular}
\end{table}

It worth mentioning that if we add other matter sources, situation with bounding $\Lambda$-term only improves in the sense that constraints becomes tighter.
Indeed, upper bound originates from the total energy density and in case of two components -- Skyrme and $\Lambda$-term, one can say that ``all what is not
Skyrme is $\Lambda$-term''; but if there are other components, $\Lambda$ is bounded by ``new'' energy budget with all components taken into account.
Lower bound on $\Lambda$ just scales with upper bound and unlikely to be highly affected by additional sources -- if they are isotropic, they ``help''
isotropize model in the similar way $\Lambda$-term do. Needless to say that the same is true for Kantowski-Sachs model as well.

\section{Discussion and conclusions}
We have studied the cosmological dynamics for the Skyrme matter field for two anisotropic cosmologies namely Kantowski-Sachs and Bianchi-I. 
To the bests of authors' knowledge this is the first detailed analysis of the cosmological consequences of gravitating Skyrme matter.
Due to the high 
nonlinearity of the equations of motion it is not possible to find simple analytic solutions. The basic features of the cosmological dynamics can however be 
studied  performing a  numerical analysis. Despite the fact that these two cosmologies are topologically very different they share several basic common 
features. Indeed in order for solutions to exist and to be free of finite time future singularities  bounds must be set on the value of the cosmological 
constant and also on the Skyrme couplings.
 
 In order to get a physically sensible scenario the cosmological constant must be strictly positive (i.e. non-zero). The upper bound on the cosmological 
 constant has a physical interpretation in terms of the phase transition from quark-gluon plasma to a confined (colorless) theory.  Indeed there are physical 
 fields whose cosmological evolution begin below the energy scale of 1 GeV since they only appear as low energy degrees of freedom of
QCD after the confinement phase transition. Hence, the initial data for these fields must be set when the total energy density is of that order. Then,
upper and lower bounds on $\Lambda$
naturally appear, as it has been explained before, when the Einstein equations are solved (numerically). Thus, the key point is just to identify a set of
physical fields which appear as source terms in Einstein equations late enough to make the upper bound on $\Lambda$ as tight as possible. Within the
Standard Model of particle physics, the ``latest'' fields to come into play are baryons and pions and so we have used the Skyrme model (altough a description
in terms of an effective perfect fluid would work as well, as long as the initial conditions are set at the GeV scale).

Even if this physical interpretation is very natural and rather simple curiosly it has never been mentioned in literature. This analysis actually may offers 
a change of paradigm on the problem to find bounds on $\Lambda$ since, unlike what is usually expected, the tightest bounds arise from
the lowest energy hadronic particles. It is also apparent that a perturbative treatment is unsuitable since the bounds depend in a non-analytic way on the
Skyrme coupling.
The same procedure could be repeated for the phase transitions (such as the ones predicted in Grand Unified Theories) prior to the one considered in present
paper -- in that case we would have qualitatively the same behavior but with higher upper bounds on $\Lambda$. In order for the Universe to enter a new
phase dominated by some lower-energy matter fields one must ensure the absence of FTFS and existence of solutions in previous phase. The energy scale
considered in this paper
corresponds to the last step in the hierarhy of bounds on $\Lambda$ so that if the upper bound on $\Lambda$ is not satisfied then the confinement phase transition would
not had even occured.

Once the value of the cosmological constant has been fixed by the initial conditions also bounds on the Skyrme couplings appear. These bounds are related to
the peculiar structure of the Skyrme action. This can be seen by the fact that similar bounds on 
the couplings arise for both cosmological models. This suggests that this behavior is intrinsic to the coupling of the Skyrme field to General Relativity 
rather than some specific cosmological {\it ansatz}. 

It is also worth mentioning that the actual upper bound is set not upon $\lambda$, $K$ and $\Lambda$ separately but, as one can see from (\ref{lambda0}) and 
(\ref{lambda0_bi}) on their linear function:

\begin{equation}
\Lambda_0 = a - bK - cK\lambda
\end{equation}

\noindent (in Bianchi-I case $c=0$); that way $K_{cr}$ and $\lambda_{cr}$ are intersections of the resulting plane with $K$ and $\lambda$ axis. This way
the resulting classification of regimes (Tables \ref{table_ks} and \ref{table_bi}) is obtained for fixed $K$ and $\lambda$ satisfing $\lambda < \lambda_{cr}$
and $K < K_{cr}$.

If one perfroms a perturbative analysis in the initial data starting from a cosmological solution of the Einstein equation and turns on ``slowly'' the Skyrme couplings then the most interesting 
bounds we found would be lost (since they are not analytic in the Skyrme couplings). 
As an alternative, one could perfrom a perturbative analysis in the initial data starting from a time-dependent solution of the Skyrme field equations turning on ``slowly'' the Newton constant. 
This alternative approach is complicated by the lack of suitable exact time-dependent solutions of the Skyrme field equations. We leave these two points for a future investigation.

\acknowledgments

This work has been funded by the FONDECYT grants 1120352, 1110167, 3130599.
The Centro de Estudios Cient\'{\i}ficos (CECs) is funded by the Chilean
Government through the Centers of Excellence Base Financing Program of Conicyt.

\end{document}